\documentclass[12pt]{article}
\usepackage{graphicx}
\usepackage{amsfonts}
\usepackage{latexsym}
\usepackage{amsmath}

\makeatletter
\@addtoreset{figure}{section}
\def\thefigure{\thesection.\@arabic\c@figure}
\@addtoreset{table}{section}
\def\thetable{\thesection.\@arabic\c@table}

\def\@sect#1#2#3#4#5#6[#7]#8{\ifnum #2>\c@secnumdepth
     \def\@svsec{}\else
     \refstepcounter{#1}\edef\@svsec{\csname the#1\endcsname.\hskip .75em
}\fi
     \@tempskipa #5\relax
      \ifdim \@tempskipa>\z@
        \begingroup #6\relax
          \@hangfrom{\hskip #3\relax\@svsec}{\interlinepenalty \@M #8\par}%
        \endgroup
       \csname #1mark\endcsname{#7}\addcontentsline
         {toc}{#1}{\ifnum #2>\c@secnumdepth \else
                      \protect\numberline{\csname the#1\endcsname}\fi
                    #7}\else
        \def\@svsechd{#6\hskip #3\@svsec #8\csname #1mark\endcsname
                      {#7}\addcontentsline
                           {toc}{#1}{\ifnum #2>\c@secnumdepth \else
                             \protect\numberline{\csname the#1\endcsname}\fi
                       #7}}\fi
     \@xsect{#5}}
\def\@begintheorem#1#2{\it \trivlist \item[\hskip \labelsep{\bf #1\ #2.}]}
\def\section{\@startsection {section}{1}{\z@}{-3.5ex plus -1ex minus
 -.2ex}{2.3ex plus .2ex}{\normalsize\bf}}

\begin{document}

\title{Efficient Parallel Simulations of \\
Asynchronous Cellular Arrays}
\date{} 
\maketitle

\begin{center}
\author{Boris D. Lubachevsky\\
{\em bdl@bell-labs.com}\\
Bell Laboratories\\
600 Mountain Avenue\\
Murray Hill, New Jersey}
\end{center}

\setlength{\baselineskip}{0.995\baselineskip}
\normalsize
\vspace{0.5\baselineskip}
\vspace{1.5\baselineskip}

\begin{abstract}
A definition for a class of asynchronous cellular arrays
is proposed.
An example of such asynchrony would be
independent Poisson arrivals of cell iterations.
The Ising model in the continuous time formulation of Glauber
falls into this class.
Also proposed are efficient parallel algorithms for
simulating these asynchronous cellular arrays.
In the algorithms, one or several cells are assigned to a processing
element (PE),
local times for different PEs can be different.
Although the standard serial algorithm by
Metropolis, Rosenbluth, Rosenbluth, Teller, and Teller
can simulate such arrays,
it is usually
believed to be without an efficient parallel counterpart.
However, the proposed parallel algorithms
contradict this belief
proving to be
both efficient
and able to perform the same task
as the standard algorithm.
The results of experiments with the new algorithms
are encouraging:
the speed-up is greater than 16
using 25 PEs on a shared memory MIMD
bus computer,
and greater than 1900
using $2^{14}$ PEs on a
SIMD computer.
The algorithm by
Bortz, Kalos, and Lebowitz
can be incorporated
in the proposed parallel algorithms,
further contributing to speed-up.
\end{abstract}
\section{Introduction}\label{sec:intro}
\hspace*{\parindent} 
Simulation is inevitable
in studying the evolution
of complex cellular systems.
Large cellular array simulations might require long runs
on a serial computer.
Parallel processing,
wherein each cell or a group of cells
is hosted by a separate processing element (PE),
is a feasible method to speed up the runs.
The strategy of a parallel simulation
should depend on whether the
simulated system is synchronous
or asynchronous.

A {\em synchronous} system
evolves in discrete time $t=0,1,2,...$.
The state of a cell at $t+1$
is determined by the state of the cell and its neighbors
at $t$
and may explicitly depend
on $t$ and the result of a random experiment.
  
An obvious and correct way to simulate
the system synchrony using a parallel processor
is simply to mimic it by the executional synchrony.
The simulation is arranged in rounds
with
one round corresponding to one time step
and with
no PE processing state changes of its cells for time $t+1$
before all PEs have processed state changes of their cells
for time $t$.
  
An {\em asynchronous} system evolves in continuous time.
State changes at different cells occur
asynchronously at unpredictable random times.
Here two questions should be answered:
(A) How to specify the asynchrony precisely?
and (B) How to carry out the parallel simulations
for the specified asynchrony?
  
Unlike the synchronous case,
simple mimicry does not work well
in the asynchronous case.
When Geman and Geman \cite{GG}, for example,
employ executional {\em physical} asynchrony
(introduced by different speeds of different PEs)
to mimic the model asynchrony,
the simulation becomes irreproducible
with its results depending on executional timing.
Such dependence may be tolerable in tasks
other than simulation
(\cite{GG} describes one such task,
another example is given in \cite{LM}).
In the task of simulation, however, it is
a serious shortcoming as seen in the following example.

Suppose  a simulationist,
after observing the results of a program run,
wishes to look closer at a certain phenomenon
and inserts an additional `print' statement
into the code.
As a result of the insertion,
the executional timing changes
and the phenomenon under investigation vanishes.

Ingerson and Buvel \cite{INBUV} and
Hofmann \cite{HOF}
propose various reproducible
computational
procedures to simulate asynchronies
in cellular arrays.
However no uniform principle has been proposed,
and no special attention to developing
parallel algorithms has been paid.
It has been observed that the
resulting cellular patterns
may depend on the computational
procedure \cite{INBUV}.

Two main results of this paper are:
(I) a definition
of a natural  class of asynchronies
that can be associated with
cellular arrays
and
(II) efficient parallel algorithms to simulate
systems in this class.
The following properties specify
the {\em Poisson asynchrony},
a most common
member in the introduced class:
\\
\\   
$~~~$Arrivals
for a particular cell
form a Poisson point process.
\\  
$~~~$Arrivals processes for different cells are independent.
\\   
$~~~$The arrival rate
is the same, say $\lambda$,
for each cell.
\\
$~~~$When there is an arrival,
the state of the cell
instantaneously changes;
the new state is computed
based on the states of the cell and its neighbors
just before the change
(in the same manner as in the synchronous model).
The new state may be equal to the old one.
\\
$~~~$The time of arrival
and a random experiment may be involved in the computation.
\\

A familiar example of a cellular system with the Poisson asynchrony
is the Ising model \cite{ISING}
in the continuous time formulation of Glauber \cite{GL}.
In this model
a cell configuration is defined
by the spin variables $s(c)=\pm 1$
specified at the cells $c$ of a two or three dimensional
array.
When there is an arrival at a cell $c$,
the spin $s(c)$ is changed to $-s(c)$ with probability $p$.
With probability $1~-~p$,
the spin $s(c)$ remains unchanged.
The probability $p$
is determined
using the values of $s(c)$ and neighbors $s(c')$ just before
the update time.

It is instructive to review the
computational procedures for Ising simulations.
First, the Ising simulationists realized that the standard procedure by
Metropolis, Rosenbluth, Rosenbluth, Teller, and Teller \cite{MRRTT}
could be applied.
In this procedure, the evolution of the configuration
is simulated as a sequence of one-spin updates:
Given a configuration,
define the next configuration by choosing a cell $c$
uniformly at random and changing or not changing the spin
$s(c)$ to $-s(c)$ as required.
In the original standard procedure time is discrete.
Time continuity could have been simply introduced
by letting
the consecutive arrivals form
the Poisson process with rate $\lambda N$,
where $N$ is the total number
of spins (cells) in the system.
  
The problem of long simulation runs became immediately apparent.
Bortz, Kalos, and Lebowitz \cite{BKL}
developed a serial algorithm (the BKL algorithm)
which avoids processing unsuccessful
state change attempts,
and reported up to a 10-fold speed-up over the 
straight-forward implementation of the
standard model.
Ogielski \cite{OGI} built special purpose hardware
for speeding up the processing.

The BKL algorithm is serial.
Attempts were made 
to speed up the Ising simulation by parallel
computations 
(Friedberg and Cameron \cite{FC}, Creutz \cite{CR}).
However, in these computations the original Markov chain
of the continuous time Ising model
was modified to satisfy the computational procedure.
The modifications do not affect the equilibrium
behavior of the chain,
and as such are acceptable
if one studies only the equilibrium.
In the cellular models however,
the transient behavior is also of interest,
and no model revision should be done.

This paper presents
efficient methods for parallel simulation
of the continuous time asynchronous cellular arrays
without changing the model or type of asynchrony in favor
of the computational procedure.
The methods
promise unlimited
speed-up when the array and the parallel
computer are sufficiently large.
For the Poisson asynchrony case, 
it is also shown how 
the BKL algorithm can be incorporated,
further contributing to speed-up.

For the Ising model,
presented algorithms can be viewed
as exact parallel counterparts
to the standard algorithm by Metropolis et al. 
The latter has been known and
believed to be inherently serial since 1953.
Yet, the presented algorithms are parallel, efficient, and fairly simple.
The ``conceptual level'' codes are rather short
(see Figures~\ref{fig:a1c1pe}, 
\ref{fig:s1c1pe}, 
\ref{fig:amcgen}, 
\ref{fig:amcpoi}, 
and
\ref{fig:genout}, 
).
An implementation in a real programming language 
given in the Appendix
is longer, of course,
but still rather simple.

This paper is organized as follows:
Section \ref{sec:model} presents 
a class of asynchronies
and a comparison with other published proposals.
Then Section \ref{sec:algo} describes the new algorithms 
on the conceptual level.
While the presented algorithms are simple,
there is no simple theory which predicts
speed-up of these algorithms for 
cellular arrays and parallel processors
of large sizes.
Section \ref{sec:perf} contains a simplified computational
procedure which predicts speed-ups faster than it takes
to run an actual parallel program.
The predictions made by this
procedure are compared with actual runs
and appear to be rather accurate.
The procedure predicts speed-up of more than 8000
for the simulation of $10^5 \times 10^5$ 
Poisson asynchronous cellular array in parallel
by $10^4$ PEs.
Actual speed-ups obtained thus far were:
more than 16 on 25 PEs of the
Balance (TM)
computer and more than 1900
on $2^{14}$ PEs of the Connection Machine (R).
\footnotetext{ 
Connection Machine is a registered trademark of Thinking Machines Corporation
\\
Balance is a trademark of Sequent Computer Systems, Inc.}

\section{Model}\label{sec:model}
\hspace*{\parindent} 
Time $t$ is continuous.
Each cell $c$ has a state $s=s(c)$.
At random times, a cell is granted a chance
to change the state.
The changes, if they occur,
are instantaneous events.
Random attempts to change the state of a cell
are independent of
similar attempts for other cells.

The general model consists
of two functions: 
{\em time\_of\_next\_arrival ()}
and {\em next\_state ()}.
They are defined as follows:
given the old state of the cell
and the states of the neighbors just before time $t$,
$s_{t-0} (neighbors (c))$,
the next\_state 
$s(c)=s_t (c)$ is
\begin{equation}
\label{newst}
s_t (c) = next\_state~(c,~s_{t-0} (neighbors(c)),~ \omega ,~t ),
\end{equation}
where the possibility 
$s_t (c)=s_{t-0} (c)$ is not excluded;
and
the time $next\_t$ of the next arrival 
is 
\begin{equation}
\label{newti}
next\_t = time\_of\_next\_arrival~(c, s_{t-0} (neighbors(c)),~ \omega ,~t),
\end{equation}
where always $next\_t ~ > ~ t$.

In \eqref{newst} and \eqref{newti}, 
$\omega$ denotes the result of a random experiment, 
e.g., coin tossing,
$s (neighbors(c))$ denotes the indexed set of states 
of all the neighbors of $c$ including $c$ itself.
Thus,
if $neighbors(c)=\{ c, c_1, c_2, c_3, c_4\}$,
then 
$s(neighbors(c)) = 
(s(c), s(c_1 ), s(c_2 ), s(c_3 ), s(c_4 ))$.
Subscript $t-0$ expresses the idea of `just before $t$',
e.g.,
$a_{t-0} ( \tau ) = lim_{\tau \rightarrow t,~\tau < t} ~ a( \tau )$.
According to \eqref{newst}, the value of $s(c)$
instantaneously changes at time $t$ 
from $s_{t-0} (c)$ to $s_t (c)$.
At time $t$, the value of $s(c)$ is already new.
The `just before' feature resolves
a possible ambiguity
if two neighbors attempt to change their states
at the same simulated time.

Compare now the class of asynchronies 
defined by \eqref{newti} with the ones proposed in the literature:
    
\ \ \ (A) Model 1 in \cite{INBUV} reads: 
``...the cells iterate randomly, one at a time.'' 
Let $p_c$ be the probability that cell $c$ is chosen.
Then the following choice of law \eqref{newti} yields this model
\[
time\_of\_next\_arrival~(c,~\omega ,~t)= t~-~ \frac {1} {p_c}   \ln  r(c,t, \omega ),
\]
where $r(c, t, \omega )$ is a random number uniformly distributed on (0,1),
and $\ln$ is the natural logarithm,
$\ln (x) = {\log}_e (x)$.
For $p_{c_1} = p_{c_2} = ... = \lambda$,
the asynchrony was called the {\em Poisson asynchrony} in Section~\ref{sec:intro};
it coincides with the one defined
by the standard model \cite{MRRTT},
and by Glauber's model \cite{GL} for the Ising spin simulations.
   
\ \ \ (B) Model 2 in \cite{INBUV} assigns
``each cell a period according to a Gaussian distribution...
The cells iterate one at a time each having its own definite
period.''
While it is not quite clear from \cite{INBUV} 
what is meant by a ``definite period''
(is it fixed for a cell over a simulation run?),
the following choice of law \eqref{newti} yields this model
in a liberal interpretation:
\[
time\_of\_next\_arrival~(c,~\omega ,~t)= t~+~ {P_c}^{-1} (r( \omega )),
\]
where $P^{-1} (y)=x$ if $P(x)=y$,
and $P_c (x)$ is the cumulative function for the
Gaussian probability distribution
with mean $m_c~>~0$ 
and variance ${\sigma_c}^2$.
The probability of
$next\_t < t$ is small when $\sigma < < m$
and is ignored in \cite{INBUV}
if this interpretation is meant.
In a less liberal interpretation,
$\sigma_c \equiv 0$ for all $c$,
and $m_c$ is itself 
random and distributed according to the Gaussian law.
This case is even easier to represent in terms of 
model \eqref{newti} than the previous one:
$time\_of\_next\_arrival^ (c,~ \omega ,~t)= t + m_c ( \omega )$.
   
\ \ \ (3) Model \eqref{newti} trivially extends to a synchronous simulation,
where the initial state changes arrive at time 0 and 
then always $next\_t - t$ is identical to 1.
The first model in \cite{HOF} is
``to choose a number of cells at random and change
only their values before continuing.''
This is a variant of synchronous simulation;
it is substantially different from both models (A) and (B) above.
In (A) and (B),
the probability is 1 that
no two neighbors attempt to change their states at the same time.
In contrast, in this model many neighboring cells 
are simultaneously changing their values.
How the cells are chosen for update
is not precisely specified in \cite{HOF}.
One way to choose the cells is to assign a probability weight
$p_c$ for cell $c$, $c=1,2,...,N$,
and to attempt to update cell $c$ 
at each iteration, 
with probability $p_c$,
independent of any other decision.
Such a method 
conforms with the law \eqref{newti}
because the method is local:
a cell does not need to know 
what is happening at distant cells.
The second model in \cite{HOF}
changes states of a
fixed number $A$ of randomly chosen cells
at each iteration.
If $A > 1$,
this method is not local
and does not conform with the law \eqref{newti}.

\section{Algorithms}\label{sec:algo}
\hspace*{\parindent} 
{\bf Elimination of $\omega$}.
Deterministic computers 
represent randomness by using
pseudo-random number generators.
Thus, equations \eqref{newst} and \eqref{newti} are substituted
in the computation by equations
\begin{equation}
\label{news0t}
s_t (c) = next\_state~(c,~s_{t-0} (neighbors(c)),~t ),
\end{equation}
and
\begin{equation}
\label{newt0i}
next\_t = time\_of\_next\_arrival~(c, s_{t-0} (neighbors(c)),~t),
\end{equation}
respectively,
which do not contain the parameter of randomness $\omega$.

This elimination of $\omega$ symbolizes
an obvious but important 
difference between the simulated system and the simulator:
In the simulated system, 
the observer, being a part of the system,
does not know in advance
the time of the next arrival.
In contrast, the simulationist who is,
of course, not a part of the simulated system,  
can know the time of the next arrival
before the next arrival is processed.

For example,
it is not known in advance when the next event
from a Poisson stream arrives.
However, in the simulation,
the time $next\_t$ of the next arrival
is obtained in a deterministic manner, 
given the time $t$ of the previous arrival:
\begin{equation}
\label{newt1i}
next\_t = t ~-~ \frac{1}{\lambda} {\log}_e ( r(n(t))),
\end{equation}
where $\lambda$ is the rate,
$r(n)$ is the $n$-th pseudo-random number in the sequence
uniformly distributed on $(0,1)$,
and $n(t)$ is the invocation counter. 
Thus, after the previous arrival is processed,
the time of the next arrival is already known.
If needed, the entire sequence of arrivals
can be precomputed and stored in a table for later
use in the simulation,
so that all future arrival times 
would be known in advance.
\\

{\bf Asynchronous one-cell-per-one-PE algorithm}.
The algorithm in Figure~\ref{fig:a1c1pe}
is the shortest of those presented in this paper.

To understand this code, 
imagine a parallel computer which consists
of a number of PEs running concurrently.
One PE is assigned to simulate one cell.
The PE which is assigned to simulate cell $c_0$,
PE$c_0$, executes the code in Figure~\ref{fig:a1c1pe} with $c=c_0$.
The PEs are interconnected by the network
which matches the topology of the cellular array.
A PE can receive information from its neighbors.
PE$c$ maintains state $s(c)$
and local simulated time $t(c)$.
Variables $t(c)$ and $s(c)$ are visible
(accessible for reading only) by the neighbors of $c$.
Time $t(c)$ has no connection with the physical
time in which the parallel computer runs the program
except that $t(c)$ may not decrease 
when the physical time increases.
At a given physical instance of simulation, 
different cells $c$ may have different values of $t(c)$.
Value $end\_time$ is a constant which is known to all PEs.

The algorithm in Figure~\ref{fig:a1c1pe}
is very asynchronous:
different PEs can
execute different steps
concurrently
and can run
at different speeds.
A statement `wait\_until~~{\em condition}', 
like the one at Step 2 in Figure~\ref{fig:a1c1pe},
does not imply
that the {\em condition} must be detected immediately after
it occurs.
To detect the {\em condition} 
at Step 2
involving local times 
of neighbors
a PE can poll 
its neighbors
one at a time,
in any order, 
with arbitrary delays,
and
without any respect to 
what these PEs are doing meanwhile.
\\
\begin{figure}
\centering
\fbox{
\begin{minipage} {12.8cm}
\begin{enumerate}
\item while $t(c)~<~end\_time$\\
\hspace*{0.2in}
\{
\item~~~~~~wait\_until $t(c)~\leq~ \min_{c'~\in~neighbors(c)} t(c')$ ;
\item~~~~~~$s(c)~\leftarrow~ next\_state~(c,~ s (neighbors (c)),~t(c))$ ;
\item~~~~~~$t(c)~\leftarrow~time\_of\_next\_arrival~(c,~s (neighbors (c)),~t(c))$\\
\hspace*{0.2in}
\}
\end{enumerate}

\end{minipage}}
\caption{Asynchronous one-cell-per-one-PE algorithm}
\label{fig:a1c1pe}
\end{figure}
Despite being seemingly almost chaotic,
the algorithm in Figure~\ref{fig:a1c1pe}
is free from deadlock.
Moreover, it
produces a unique simulated trajectory
which is independent of executional
timing, 
provided that:
\\

(i) 
for the same cell, the pseudo-random sequence is always the same,
\\

(ii) no two neighboring arrival times are equal.
\\

Freedom from deadlock follows from the fact that the cell, 
whose local time is minimal over the entire array,
is always able to make progress.
(This guaranteed worst case performance,
is substantially exceeded
in an average case.
See Section~\ref{sec:perf}.)

The uniqueness of the trajectory can be seen as follows.
By (ii), 
a cell $c$ passes the test at Step 2 only if its local time $t(c)$
is smaller than the local time $t(c')$
of any its neighbor $c'$.
If this is the case, then no neighbor $c'$
is able to pass the test at Step 2 before
$c$ changes its time at Step 4.
This means that processing of the update by $c$ is safe:
no neighbor changes its state or time before $c$ completes
the processing.
By (i), functions $next\_state( )$ and $time\_of\_next\_arrival( )$
are independent of the run.
Therefore, 
in each program run,
no matter what the neighbors of $c$
are doing or trying to do,
the next arrival time and state for $c$ are always the same.

It is now clear why assumption (ii) is needed.
If (ii) is violated by two cells $c$ and $c'$ which are neighbors,
then the algorithm in Figure~\ref{fig:a1c1pe}
does not exclude concurrent updating by $c$ and $c'$.
Such concurrent updating
introduces an indeterminism
and inconsistency.
A scenario
of the inconsistency
can be as follows:
at Step 3
the {\em old} value of $s(c')$ is used
to update state $s(c)$,
but 
immediately following Step 4 uses the
{\em new} value of $s(c')$ 
to update time $t(c)$. 

In practice, the algorithm in Figure~\ref{fig:a1c1pe} is safe,
when $next\_t(c)-t(c)$ for different $c$ are independent 
random samples from a distribution with a continuous density,
like an exponential distribution.
In this case, (ii) holds with probability 1.
Unless the pseudo-random number generators are faulty,
one may imagine only one reason for violating (ii):
finite precision of computer representation of real numbers.
\\

{\bf Synchronous one-cell-per-one-PE algorithm}.
If (ii) can be violated with a positive probability
(if $t$ takes on only integer values,
for example),
then the errors might not be tolerable.
In this case the synchronous algorithm in
Figure~\ref{fig:s1c1pe} should be used.

Observe that while the algorithm in Figure~\ref{fig:s1c1pe} 
is synchronous,
it is able to simulate correctly
both synchronous and asynchronous systems.
Two main additions 
in the algorithm in Figure~\ref{fig:s1c1pe}
are:
private variables $new\_s$ and $new\_t$ 
for temporal storage of updated $s$ and $t$,
and synchronization barriers `synchronize'.
When a PE hits a `synchronize' statement it must wait until
all the other PEs hit a `synchronize' statement;
then it may resume.
Two dummy synchronizations at Steps 9 and 10 are executed
by idling PEs in order to match synchronizations
at Steps 5 and 8 executed by non-idling PEs.

When (ii) is violated,
the synchronous algorithm avoids the ambiguity and indeterminism
(which in this case are possible in the asynchronous algorithm)
as follows:
in processing concurrent updates of two neighbors $c$ and $c'$ 
for the same simulated time $t=t(c)=t(c')$,
first, $c$ and $c'$ read states $s_{t-0}$ and times $t$ of each other
and compute their private $new\_s$'s and $new\_t$ 
(Steps 3 and 4 in Figure~\ref{fig:s1c1pe});
then, after the synchronization barrier at Step 5,
$c$ and $c'$ write their states and times at Steps 6 and 7,
thus making sure that no write 
interferes with a read.
\\
\begin{figure}
\centering
\fbox{
\begin{minipage} {14.8cm}
\begin{enumerate}
\item while $t(c)~<~end\_time$\\
\hspace*{0.2in}
\{
\item~~~~~~if $t(c)~\leq~ \min_{~c'\in neighbors(c)} ~t(c')$ then\\
\hspace*{0.2in}
~~~~~~~~\{
\item~~~~~~~~~~~~~~$new\_s ~\leftarrow~ ~next\_state~(s (neighbors (c)),~t(c))$ ;
\item~~~~~~~~~~~~~~$new\_t~\leftarrow~ ~time\_of\_next\_arrival~(c,~t(c))$ ;
\item~~~~~~~~~~~~~~synchronize;~~~~~~~~~~~~~~~~~~~~~~~~~~~~~~~~~~~~~~~~~~~~~/* barrier 1 */
\item~~~~~~~~~~~~~~$s(c)~~\leftarrow~~new\_s$;
\item~~~~~~~~~~~~~~$t(c)~~\leftarrow~~new\_t$;
\item~~~~~~~~~~~~~~synchronize~~~~~~~~~~~~~~~~~~~~~~~~~~~~~~~~~~~~~~~~~~~~~~/* barrier 2 */\\
\hspace*{0.2in}
~~~~~~~~\}\\
\hspace*{0.2in}
\ \ else~\{
\item~~~~~~~~~~~~~~synchronize;~~~~~~~~~~~~~~~~~~~~~~~~~~~~~~~~~~~~~~~~~~~~~/* barrier 1 */
\item~~~~~~~~~~~~~~synchronize~~~~~~~~~~~~~~~~~~~~~~~~~~~~~~~~~~~~~~~~~~~~~~/* barrier 2 */\\
\hspace*{0.2in}
~~~~~~~~\}
\\
\hspace*{0.1in}
\ \ \}
\end{enumerate}

\end{minipage}}
\caption{Synchronous one-cell-per-one-PE algorithm}
\label{fig:s1c1pe}
\end{figure}
\\

\begin{figure}
\centering
\includegraphics*[width=5.8in]{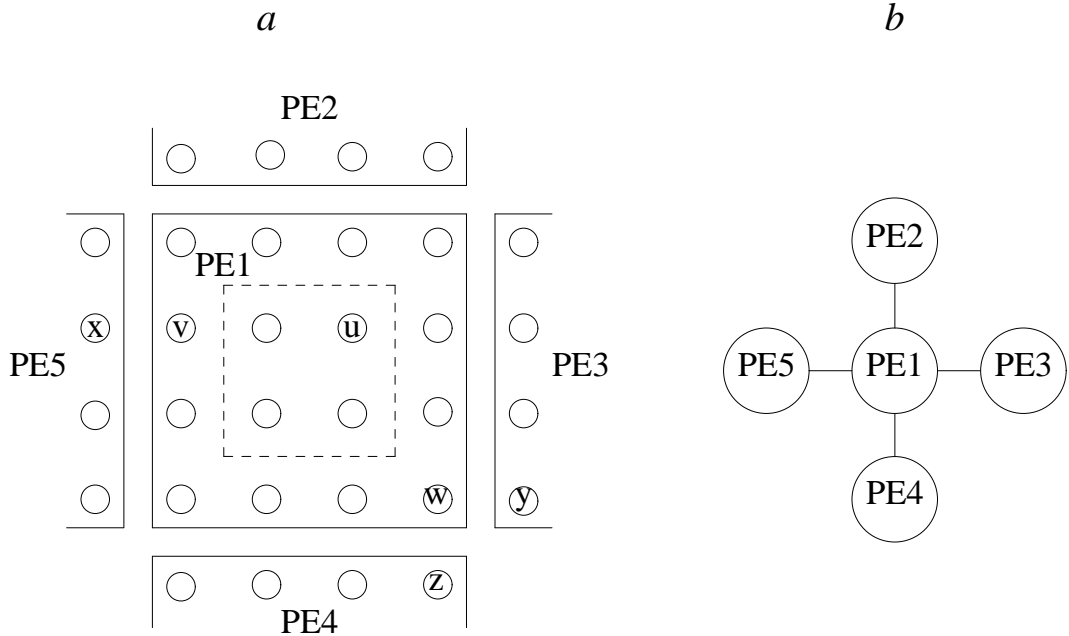}
\caption{Aggregation:
$a$) mapping of cells to PEs,
$b$) the interconnection among the PEs which supports the neighborhood
topology among the cells
}
\label{fig:aggr}
\end{figure}

{\bf Aggregation.}
In the two algorithms presented above,
one PE hosts only one cell.
Such an arrangement may be wasteful
if the communication between PEs dominates
the computation internal to a PE.
A more efficient arrangement is
to assign several cells to one PE.
For concreteness, 
consider a two-dimensional $n \times n$
array with periodic boundary conditions.
Let $n$ be a multiple of $m$ and $(n/m)^2$
PEs be available.
PE$C$ carries $m \times m$ subarray $C$,
where $C=1,2,...,(n/m)^2$.
(Capital $C$ will be used without confusion to represent
both the subarray index and the set of cells $c$ the subarray
comprises, e.g. as in $c\in C$)
A fragment of a square cellular array
in an example of such an aggregation
is represented in Figure~\ref{fig:aggr}$a$,
wherein $m=4$.

The neighbors of a cell carried by PE1 are cells carried
by PE2, PE3, PE4, or PE5.
PE1 has direct connections with these four PEs (Figure~\ref{fig:aggr}$b$).
Given cell $c$ in the subarray hosted by PE1,
one can determine with which neighboring PEs
communication is required 
in order to learn the states of the neighboring cells.
Let $W(c)$ be the set of these PEs.
Examples in Figure~\ref{fig:aggr}$a$ : $W(u)$ is empty,
$W(v)=$\{PE5\}, $W(w)=$\{PE3, PE4\}.

\begin{figure}
\centering
\fbox{
\begin{minipage} {13.8cm}
\begin{enumerate}
\item while $T(C) ~<~ end\_time$ 
\\
\hspace*{0.2in}
\{
\item~~~~~~select a cell $c$ in the subarray $C$ such that\\
\hspace*{0.2in}
~$t(c)= \min_{~c' \in C} ~t(c')$ and assign $T(C)\leftarrow t(c)$;
\item~~~~~~wait\_until $T(C)~\leq~ \min_{~C' \in W(c)} ~T(C')$ ;
\item~~~~~~$s(c) \leftarrow next\_state~(c,~ s (neighbors (c)),~t(c))$ ;
\item~~~~~~$t(c) \leftarrow time\_of\_next\_arrival~(c,~s (neighbors (c)),~t(c))$\\
\hspace*{0.2in}
\}
\end{enumerate}
\end{minipage}}
\caption{Asynchronous many-cells-per-one-PE algorithm. General asynchrony}
\label{fig:amcgen}
\end{figure}
Figure~\ref{fig:amcgen} presents an aggregated variant of the algorithm
in Figure~\ref{fig:a1c1pe}.
PE$C$, which hosts
subarray $C$,
maintains the local time register $T(C)$.
PE$C_0$ simulates the evolution of its subarray
using the algorithm in Figure~\ref{fig:amcgen}
with $C=C_0$.
Each cell $c~\in~C$
is represented in the memory of PE$C$
by its current state $s(c)$ and its
next arrival time $t(c)$.
Note that unlike the one-cell-per-one-PE algorithm,
the $t(c)$ does not represent the current local time for cell $c$.
Instead, local times of all cells within subarray $C$ are the same, $T(C)$. 

\begin{figure}
\centering
\includegraphics*[width=6.2in]{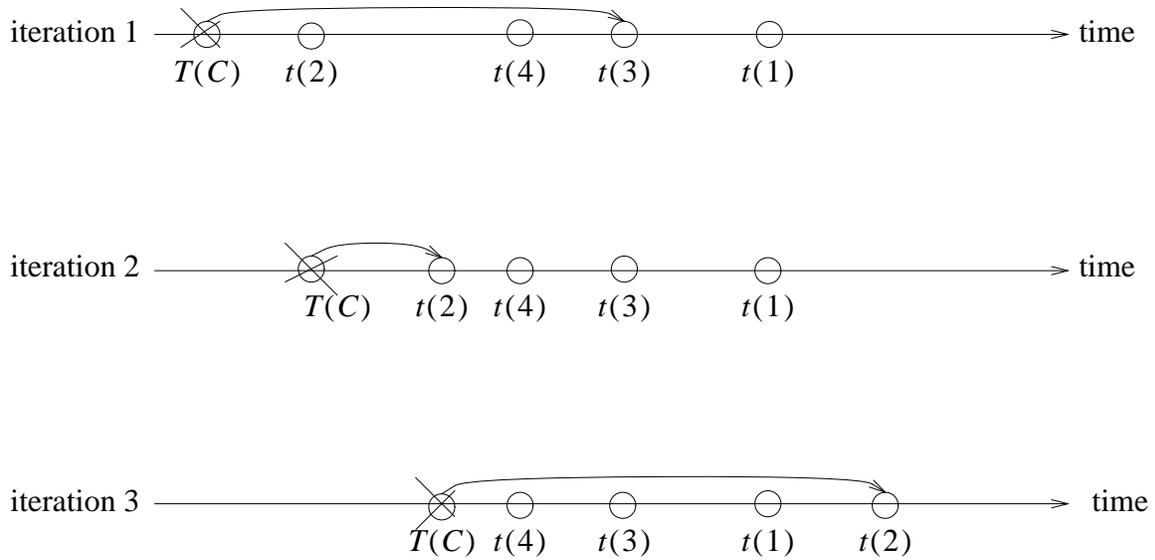}
\caption{$T(C)$ slides along a sequence of $t(c)$'s 
in successive iterations of the aggregated algorithm}
\label{fig:timlin}
\end{figure}

$T(C)$ moves from one $t(c)$ to another 
in the order of increasing value.
Three successive iterations of this algorithm
are shown in Figure~\ref{fig:timlin}, where the subarray $C$
consists of four cells: $C=\{1, 2, 3, 4 \}$.
Circles in Figure~\ref{fig:timlin} represent 
arrival points in the simulated time.
A crossed-out circle represents an arrival which
has just been processed,
i.e., Steps 3, 4, and 5 of Figure~\ref{fig:amcgen} have just been executed,
so that $T(C)$ has just taken on the value of the processed
old arrival time $t(c)$, 
while the $t(c)$ has taken on a new larger value.
This new value is pointed to by an arrow from $T(C)$ in Figure~\ref{fig:timlin}.
It is obvious that 
always $t(c)~>=~T(C)$ if $c~\in~C$.

Local times $T(C)$ 
maintained by different PE$C$ might be different.
A wait at Step 3 cannot deadlock
the execution
since the PE$C$ whose $T(C)$ is the minimum over
the entire cellular array is always able to make a progress.

Assuming property (ii) as above,
the algorithm 
correctly simulates the history
of updates.
The following example may serve as an informal proof of this statement.
Suppose PE1 is currently updating the state of cell $v$ 
(see Figure~\ref{fig:aggr}$a$)
and its local time is
$T_1$.
Since $W(v)=\{ PE5\}$,
this update is possible
because the local time of PE5, $T_5$, is currently
larger than $T_1$.
At present,
PE1 receives the state
of $x$ from PE5
in order to perform the update.
This state 
is in time $T_5$,
i.e., in the future with respect to local time $T_1$.
However, the update is correct,
since 
the state of $x$ was the same at time $T_1$, 
as it is at time $T_5$.

Indeed, suppose  
the state of $x$ were to be changed 
at simulated local time $T$, 
$T_1 <  T <  T_5$.
At the moment when this change would have been processed by PE5,
the local time of PE1 would have been larger than $T$,
and $T$ would have been the local time of PE5.
After this processing has supposedly taken place,
the local time of PE1 should not decrease.
Yet at the present it is $T_1$,
which is smaller that $T$.
This contradiction proves that the state
of $x$ cannot in fact change
in the interval ($T_1, T_5$).

In the example in Figure~\ref{fig:timlin},
only one $t(c)$ supplies $\min_{~c' \in C} t(c')$.
However, the algorithm in Figure~\ref{fig:amcgen}
at Step 2 commands to select {\em a} cell
not {\em the} cell.
This covers the unlikely situation of several cells having the same
minimum time.
If $next\_t(c) - t(c)$ for different $c$ are independent 
random samples from a distribution with a continuous density,
this case occurs with the probability zero.
On the other hand, 
if several cells can, with positive probability,
update simultaneously,
a synchronous version of the aggregated algorithm should be used instead.
To eliminate indeterminism and inconsistency,
the latter would use
synchronization and intermediate storage
techniques.
These techniques were demonstrated in the algorithm in Figure~\ref{fig:s1c1pe}
and their discussion is not repeated here.

\begin{figure}
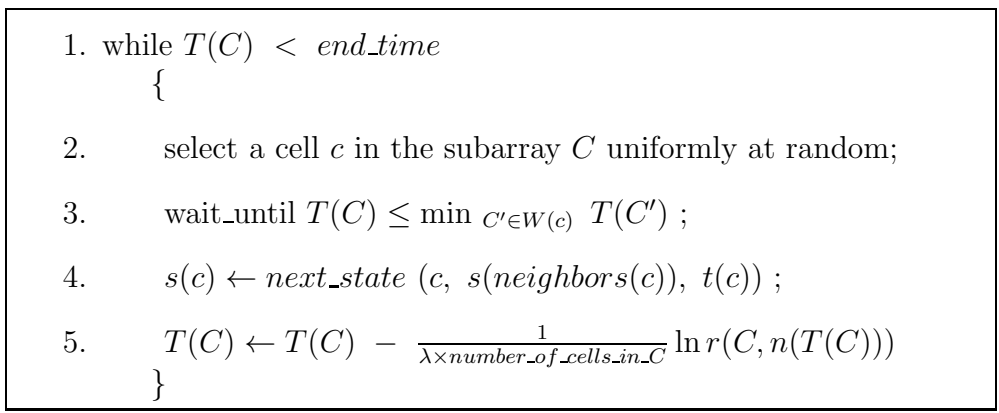

\centering
\fbox{
\begin{minipage} {12.8cm}
\begin{enumerate}
\item while $T(C)~<~end\_time$ \\
\hspace*{0.2in}
\{
\item~~~~~~select a cell $c$ in the subarray $C$ uniformly at random;
\item~~~~~~wait\_until $T(C) \leq \min_{~C' \in W(c)} ~T(C')$ ;
\item~~~~~~$s(c) \leftarrow next\_state~(c,~ s (neighbors (c)),~t(c))$ ;
\item~~~~~~$T(C) \leftarrow T(C)~-~ \frac{1}{\lambda \times number\_of\_cells\_in\_C} \ln r(C,n(T(C)))$\\
\hspace*{0.2in}
\}
\end{enumerate}

\end{minipage}}
\caption{Asynchronous many-cells-per-one-PE algorithm. Poisson asynchrony}
\label{fig:amcpoi}
\end{figure}

For an important special case of 
{\bf Poisson asynchrony in the aggregated algorithm},
the algorithm of Figure~\ref{fig:amcgen}
is rewritten in Figure~\ref{fig:amcpoi}.
This specialization capitalizes on the 
additive property of Poisson streams,
specifically, on the fact 
that sum of $k$ independent Poisson streams
with rate $\lambda$ each
is a Poisson stream with rate $\lambda k$.
In the algorithm,
$k=number\_of\_cells\_in\_C$;
this $k$ is equal to $m^2$ in the special case of partitioning
into $m \times m$ subarrays.
Unlike the general algorithm of Figure~\ref{fig:amcgen},
in the specialization in Figure~\ref{fig:amcpoi}
neither individual streams
for different cells
are maintained,
nor future arrivals $t(c)$ for cells are 
individually computed.
Instead, a single cumulative stream is simulated
and cells are delegated randomly
to meet these arrivals.

At Step 5 in Figure~\ref{fig:amcpoi},
$r(C, n(T(C)))$ is an $n(T(C))$-th pseudo-random
number in the sequence uniformly distributed in (0,1).
It follows from the notation
that each PE has its own sequence.
If this sequence is independent of the 
run (which is condition (i) above) and
if updates for neighboring cells never coincide in time
(which is condition (ii) above), then this algorithm produces
a unique reproducible trajectory.
The same statement is also true for the algorithm in Figure~\ref{fig:amcgen}.
However, 
uniqueness provided by the algorithm in Figure~\ref{fig:amcpoi}
is weaker than the one provided by the algorithm in Figure~\ref{fig:amcgen}: 
if the same array is partitioned differently and/or executed
with different number of PEs,
a trajectory produced by the algorithm in Figure~\ref{fig:amcpoi}
may change;
however, a trajectory produced by the algorithm in Figure~\ref{fig:amcgen}
is invariant for
such changes given that each cell $c$ uses its own
fixed
pseudo-random sequence.
\\

{\bf Efficiency of aggregated algorithms}.
Both many-cells-per-one-PE algorithms 
in Figure~\ref{fig:amcgen} and Figure~\ref{fig:amcpoi}
are more efficient than the
one-cell-per-one-PE counterparts
in Figure~\ref{fig:a1c1pe} and Figure~\ref{fig:s1c1pe}.
This additional efficiency
can be explained in the example
of the square array, as follows:
In the algorithms
in Figure~\ref{fig:a1c1pe} and Figure~\ref{fig:s1c1pe},
a PE may wait for its four neighbors.
However,
in the algorithms in Figure~\ref{fig:amcgen} and Figure~\ref{fig:amcpoi}, 
a PE waits for at most two neighbors.
For example, when the state of cell $w$ in Figure~\ref{fig:aggr}$a$ is updated,
PE1 might wait for PE3 and PE4.
Moreover,
for at least
$(m-2)^2$ cells $c$ out of $m^2$,
PE1 does not wait at all,
because $W(c)=\emptyset$.
The cells $c$ such that $W(c)=\emptyset$
form the dashed square in Figure~\ref{fig:aggr}$a$.

This additional efficiency becomes especially large if,
instead of set $neighbors (c)$ in the original formulation
of the model,
one uses sets 
\begin{equation}
\label{nei2}
neighbors^2 (c)~ \stackrel{\rm def}{=} ~next\_to\_nearest\_neighbors (c)
\end{equation}
or, more generally, $q$-th degree neighborhood, 
$neighbors^q (c)$.
The latter is
defined for $q~>~1$ inductively
\begin{equation}
\label{neiq}
neighbors^q (c) \stackrel{\rm def}{=} neighbors ( neighbors^{q-1} (c))
\end{equation}
where $neighbors (S)$ for a set $S$ of cells
is defined as
$neighbors (S) \stackrel{\rm def}{=}  \bigcup_{~c \in S} neighbors (c)$.

It is easy to rewrite 
the algorithms in Figure~\ref{fig:a1c1pe} and Figure~\ref{fig:s1c1pe}
for the case $q~>~1$.
The obtained codes have low efficiency however.
For example,
in the square array case,
one has
$| neighbors^q (c) | - 1=2q(q+1)$.
Thus, if $q=2$,
a cell might have to wait
for 12 cells
in order to update.
In the same example, 
if one PE carries an $m \times m$ subarray,
and $m~>~q$, then the PE waits for at most three other PEs
no matter how large the $q$ is.
Moreover,
if $m > 2q$ then in $(m-2q)^2$ cases out of $m^2$ 
the PE does not wait at all.
\\

{\bf The BKL algorithm} \cite{BKL}
was originally proposed for Ising spin simulations.
It was noticed that the probability $p$ to flip
$s(c)$ takes on only a finite (and small) number
$d$ of values $p_1 ,..., p_d$,
each corresponding to one or several
combinations of old values of $s(c)$ and neighboring spins $s(c')$.
Thus the algorithm
splits the cells into $d$
pairwise disjoint classes $\Gamma_1$, $\Gamma_2$,...$\Gamma_d$.
The rates $\lambda p_k$ of changes
(not just of the attempts to change)  
for all $c \in \Gamma_k$
are the same.
At each iteration, the BKL algorithm does the following: 
\\
\begin{quotation}
(a) Selects $\Gamma_{k_0}$ at random according to the 
weights $| \Gamma_k | p_k$, $k=1,2,...d$,
and selects a cell $c \in \Gamma_{k_0}$ uniformly at random.
\\
    
(b) Flips the state of the selected cell, 
$s(c) \leftarrow -s(c)$.
\\
    
(c) Increases the time by 
$- {\log}_e (r) /( \lambda ( \sum_{1 \leq k \leq d} | \Gamma_k |  p_k ))$,
where $r$ is a pseudo-random number uniformly distributed in (0,1).
\\
    
(d) Updates the membership in the classes.
\end{quotation}
If the asynchrony law is Poisson,
the idea of the BKL algorithm 
can be applied also to a 
deterministic update.
Here the probability $p$
of change takes on just two values:
\\
$p_1 =0$ if $next_s(c)=s(c)$,
and $p_2 =1$ if $next\_s(c)~ \neq ~s(c)$.
\\
Accordingly, there are two classes:
$\Gamma_0$, the cells which are not going to change
and
$\Gamma_1$, the cells which are going to change.
As with the original BKL algorithm,
a substantial overhead is required for maintaining an account 
of the membership in the classes (Step (d)).
The BKL algorithm is justified only if a large number of cells
are not going to change their states.
The latter is often the case.
For example,
in the Conways's synchronous {\em Game of Life}
(Gardner \cite{GAR})
large regions of
white cells ($s(c)=0$) remain
unchanged for many iterations
with very few black cells ($s(c)=1$).
One would expect similar behavior
for an asynchronous version of the 
Game of Life.

The basic BKL algorithm is serial.
To use it on a parallel computer,
an obvious idea is to run a copy of the serial BKL algorithm
in each subarray carried by a PE.
Such a procedure,
however,
causes roll-backs,
as seen in the following example:
   
Suppose PE1 is currently updating the state
of cell $v$ (Figure~\ref{fig:aggr}$a$) and its 
local time is $T_1$,
while the local time of PE5, $T_5$,
is larger than $T_1$.
Since $x$ is a nearest neighbor to $B$,
$x$'s membership might change because of $v$'s changed state.
Suppose $x$'s membership were to indeed change.
Although this change would have been in effect since time $T_1$,
PE5, which is responsible for $x$,
would learn about the change 
only at time $T_5 ~>~T_1$.
As the past of PE5 is not, therefore,
what PE5 has believed it to be,
interval [$T_1 , T_5$] must have
been simulated by PE5 incorrectly,
and must be played again.
This original roll-back might cause a cascade
of secondary roll-backs, third generation roll-backs etc.
\\

{\bf A modified BKL algorithm}
applies the original BKL procedure
only to a subset of the cells,
whereas
the procedure of the standard model is applied
to the remaining cells.
More specifically:
An additional separate class $\Gamma_0$ is defined.
Unlike other $\Gamma_k$, $k~>~0$, 
class $\Gamma_0$
always contains the same cells.
Steps (a) - (d) are performed as above
with the following modifications:
\newpage
\begin{quotation}
1) The weight of $\Gamma_0$ at step (a) is taken to be
$| \Gamma_0 |$.
\\

2) If the selected $c$ belongs to $\Gamma_0$,
then at step (b) the state of $c$ may or may not change.
The probability $p$ of change 
is determined as in the standard model.
\\

3) The time at step (c) should be increased by
$- {\log}_e (r) /( \lambda ( | \Gamma_0 |~+~\sum_{1 \leq k \leq d} | \Gamma_k |p_k ))$,
where $r=r(c,n(t))$ is a pseudo-random number uniformly distributed in $(0,1)$.\\
\end{quotation}

Now consider again the subarray
carried by PE1 in Figure~\ref{fig:aggr}$a$.
The subarray can be subdivided
into the $(m-2) \times (m-2)$ ``kernel'' square 
and the remaining boundary layer.
If first degree neighborhood, $neighbors~(c)$,
is replaced with the $q$-th degree neighborhood,
$neighbors^q (c)$,
then the kernel is the central $(m-2q) \times (m-2q)$ square,
and the boundary layer has width $q$.
In Figure~\ref{fig:aggr}$a$, the cells in the dashed square 
constitute the kernel with $q=1$.
To apply the modified BKL procedure to 
the subarray carried by PE1,
the boundary layer is declared to be
the special fixed class $\Gamma_0$.
Similar identification is done in the other subarrays.
As a result,
the fast concurrent BKL procedures 
on the kernels
are shielded from each other
by slower procedures on the layers.

The roll-back is avoided,
since state change of a cell
in a subarray does not constitute state
or membership change of a cell
in another subarray.
Unless the performance of PE1 is taken into account,
the neighbors of PE1
can not even tell whether PE1 
uses the standard or the BKL 
algorithm to update its kernel.
As the size of the subarray increases,
so does both the relative weight of the kernel
and the fraction of the fast BKL processing.
\\

{\bf Generating the output}.
Consider the task of generating cellular patterns
for specified simulated times.
A method for performing this task in a serial
simulation or a parallel simulation of a synchronous
cellular array is obvious:
as the global time reaches a specified value,
the computer outputs the states of all cells.
In an asynchronous simulation, 
the task becomes more complicated
because
there is no global time:
different PEs may have different local times
at each physical instance of simulation.

Suppose for example,
one wants to see the cellular patterns
at regular time intervals
$K_0 \Delta t,~(K_0 +1)  \Delta t,~(K_0 +2) \Delta t,...$
on a screen of a monitor attached to the computer.
Without getting too involved 
in the details of performing I/O
operations and the architecture of the parallel computer,
it would be enough to assume that a separate process
or processes are associated with the output;
these processes scan an output buffer memory space
allocated in one or several PEs or in the shared memory;
the buffer space consists of $B$ frames,
numbered 0,1,...,$B-1$,
each capable of storing a complete image of
the cellular array for one time instance.
The output processes draw
the image for time $K \Delta t$ 
on the screen
as soon as
the frame number $rem (K/B)$
(the reminder of the integer
division $K$ by $B$)
is full and the previous 
images have been shown.
Then the frame is flashed for
the next round when it will be filled
with the image for time $(K+B) \Delta t$
and so on.
\\
\begin{figure}
\centering
\fbox{
\begin{minipage} {12.8cm}
/* Initially $K=K_0$, $T(C)~<~K_0 \Delta t$ */\\
\begin{enumerate}
\item while $T(C)~<~end\_time$\\
\hspace*{0.15in}
\{
\item~~~~~~select a cell $c$ in the subarray $C$ such that\\
\hspace*{0.2in}
~$t(c)=\min_{~c' \in C} t(c')$ and assign $new\_T \leftarrow t(c)$;
\item~~~~~~while $new\_T > K \Delta t$ \\
\hspace*{0.2in}
~~~~\{
\item~~~~~~~~~~~~~wait\_until frame $rem (K/B)$ is available;
\item~~~~~~~~~~~~~store image $s(C)$ into frame $rem (K/B)$;
\item~~~~~~~~~~~~~$K \leftarrow  K+1$\\
\hspace*{0.2in}
~~~~~\};
\item~~~~~~$T(C)  \leftarrow  new\_T$;
\item~~~~~~wait\_until $T(C) \leq \min_{~C' \in W(c)} ~T(C')$ ;
\item~~~~~~$s(c) \leftarrow next\_state (c,~ s (neighbors (c)),~t(c))$ ;
\item~~~~~~$t(c) \leftarrow  time\_of\_next\_arrival~(c,~s (neighbors (c)),~ t(c))$\\
\hspace*{0.15in}
\}
\end{enumerate}
\end{minipage}}
\caption{Generating the output in the aggregated asynchronous algorithm}
\label{fig:genout}
\end{figure}

The algorithm must fill the appropriate frame
with the appropriate data as soon as 
both data and the frame become available.
The modifications that enable the asynchronous algorithm 
in Figure~\ref{fig:amcgen} to perform this task
are presented in Figure~\ref{fig:genout}.
In this algorithm,
variables $new\_T$ and $K$ are private (i.e., local to PE)
and
$\Delta t$ and $K_0$ are constants
whose values are the same for all the PEs.
Note that different PEs may 
fill different frames concurrently.
If the slowest PE is presently filling an image for time $K \Delta t$,
then the fastest PE is allowed to fill the image for
time no later than $(K + B - 1) \Delta t$.
An attempt by the fastest PE to 
fill the image for time $(K + B) \Delta t$
will be blocked at Step 4, 
until the frame 
number $rem(K/B)=rem((K + B)/B)$
becomes available.

Thus, the finiteness of the output buffer introduces
a restriction which is not present in the original algorithm
in Figure~\ref{fig:amcgen}.
According to this restriction,
the lag between concurrently processed local times
cannot exceed
a certain constant.
The exact value of the constant in each particular instance
depends on the relative
positions of the update times within the $\Delta t$-slots.
In any case, 
the constant is not smaller than
$(B-1) \Delta t$
and
not larger than
$B \Delta t$.

However, 
even with a single output buffer segment, $B=1$,
the simulation does not become time-driven.
In this case,
the concurrently processed local times might be
within a distance of
up to $\Delta t$
of each other,
whereas $\Delta t$ might be relatively large.
No precision of update time representation is lost,
although efficiency might degrade 
when both $\Delta t$ and $B$ become too small,
see Section~\ref{sec:perf}.

\section{Performance assessment: experiments and simulations}\label{sec:perf}
\hspace*{\parindent} 
Modeling and analysis of asynchronous
algorithms is a difficult theoretical problem.
Strictly speaking,
the following discussion is applicable
only to synchronous algorithms.
However, one may argue informally
that the performance of an asynchronous
algorithm is not worse than that of its synchronous counterpart, 
since expensive synchronizations are eliminated.

First, consider the synchronous algorithm in Figure~\ref{fig:s1c1pe}.
Let $N$ be the size of the array and $N_0$ be the number of
cells which passed
the test at Step 2, Figure~\ref{fig:s1c1pe}.
The ratio of useful work performed, 
to the total work expended at
the iteration is $N_0 /N$.
This ratio yields the {\em efficiency}
(or {\em utilization}) at the given iteration.
Assuming that in the serial algorithm all the work is useful,
and that the algorithm performs the same computation as its parallel
counterpart,
the speed-up of the parallel computation
is the average efficiency times the number of PEs involved.
Here the averaging is done 
with equal weights
over all the iterations.

In the general algorithms, 
$next\_t(c)$ is determined using the
states of the neighbors of $c$.
However, in the important applications,
such as an Ising model,
$next\_t(c)$ is independent of states.
The following assessment is valid only for
this special case of independence.
Here
the configuration is irrelevant
and
whether the test succeeds or not 
can be determined knowing only the times at each iteration.
This leads to a simplified model in which 
only local times are taken into account:
at an iteration,
the local time of
a cell is incremented
if the time does not exceed
the minimum of the local times of its neighbors.

A simple (serial) algorithm 
which updates only local times of cells $t(c)$
according to the rules formulated above
was exercised for different array sizes $n$
and three different dimensions:
for an $n$-element circular array,
an $n \times n$ toroidal array,
and for $n \times n \times n$ array with periodic boundary conditions.
Two types of asynchronies are tried:
the Poisson asynchrony
for which
$next\_t~-~t$ is distributed exponentially,
and the asynchrony
for which $next\_t~-~t$ is
uniformly distributed in (0,1).
In both cases,
random time increments
for different cells are independent.

\begin{figure}
\centering
\includegraphics*[width=5.8in]{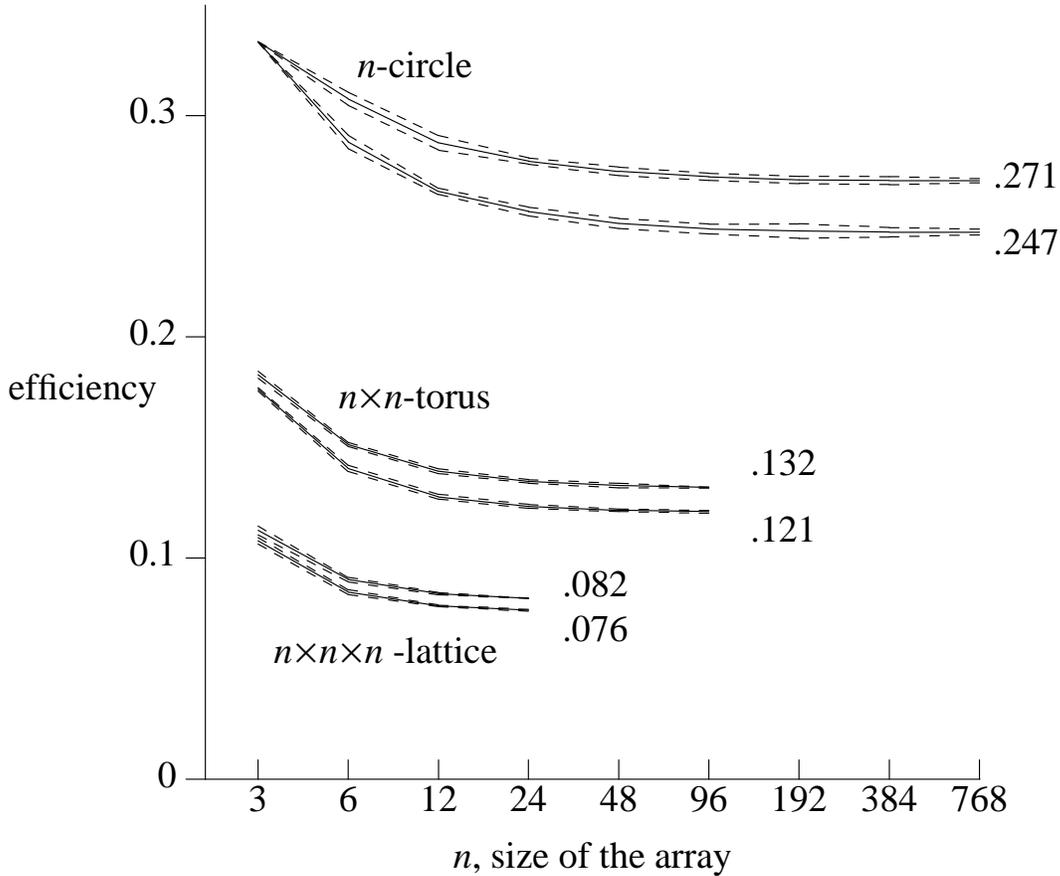}
\caption{Performance of the Ising model simulation. One-cell-per-one-PE case}
\label{fig:perf1t1}
\end{figure}

The results of these six experiments
are given in Figure~\ref{fig:perf1t1}.
Each solid line in Figure~\ref{fig:perf1t1} is enclosed between two dashed
lines. 
The latter represent
99.99\% Student's confidence intervals constructed
using several simulation runs, 
that are parametrically the same 
but fed with different pseudo-random sequences.
In Figure~\ref{fig:perf1t1}, for each array topology
there are two solids lines.
The Poisson asynchrony
always corresponds to the lower line.
The corresponding limiting values of performances
(when $n$ is large)
are also shown near the right end of each curve.
For example, the efficiency in the 
simulation of a large $n \times n$ array
with the Poisson asynchrony is about 0.121,
with the other asynchrony, it is about 0.132.

No analytical theory is available
for predicting these values
or even proving their separation from zero
when $n \rightarrow +\infty$.
It follows from Figure~\ref{fig:perf1t1} that replacing
exponential distribution of $next\_t - t$ with
the uniform distribution results in efficiency
increase
from 0.247 to 0.271 for a large $n$-circle
($n \rightarrow +\infty$).
The efficiency can be raised even more.
If $next\_t - t = r^{1/8}$,
where $r$ is distributed uniformly in (0,1),
then in the limit $n \rightarrow +\infty$,
with the Student's confidence 99.99\%,
the efficiency is $0.3388 \pm 0.0012$.
It is not known how high the efficiency
can be raised this way
(degenerated cases, like a synchronous one,
in which the efficiency is 1, are not counted).

An efficiency of 0.12 means the speed-up
of $0.12 \times N$;
for $N=2^{14}$ this comes to more than 1900.
This assessment is confirmed in an actual full scale simulation experiment
performed on $2^{14}=128 \times 128$ PEs of
a Connection~Machine~(R)
(a quarter of the full computer 
).
This SIMD computer 
appears well-suited for the synchronous execution
of the one-cell-per-one-PE algorithm in Figure~\ref{fig:s1c1pe} 
on a toroidal array,
Poisson asynchrony law.
Since an individual PE is rather slow,
it executes several thousand
instructions per second,
and its absolute speed is not very impressive:
It took roughly 1 sec. of real time
to update
all $128 \times 128$ spins
when the traffic generated by other tasks running
on the computer was small
(more precise measurement
was not available).
This includes about 
$8.3~\approx~(0.12)^{-1}$
rounds of the algorithm,
several hundred instructions of one PE per round.

The 12\% efficiency in the one-cell-per-one-PE
experiments could be greatly increased
by aggregation.
The many-cells-per-one-PE
algorithm in Figure~\ref{fig:amcpoi} is implemented
as a $C$ language parallel program for a 
Balance~(TM) computer,
which is a shared memory MIMD bus machine.
The $n \times n$ array was split into 
$m \times m$ subarrays, 
as shown in Figure~\ref{fig:aggr},
where $n$ is a multiple of $m$.
Because the computer has 30 PEs,
the experiments could be performed only with
$(n/m)^2=1, 4, 9, 16$, and 25 PEs
for different $n$ and $m$.

Along with these experiments,
a simplified model, similar to
the one-cell-per-one-PE case,
was run on a serial computer.
In this model,
quantity 
$h(C) \stackrel{\rm def}{=} \lambda T(C)$ is maintained for each PE,
$C=1,...,(n/m)^2$.
The update of $h(C)$ is arranged in rounds,
wherein each $h(C)$ is updated as follows:
\\
~~~~~~(i) with probability $p_0 =(m-2)^2 /m^2$,
PE$C$ updates $h(C)$:
\begin{equation}
\label{hC}
h(C) ~\leftarrow~ h(C) ~-~ \ln ^r (C, n(h(C))),
\end{equation}
where $r$ and $\ln$ are the same as in Step 5 in 
Figure~\ref{fig:amcpoi}.
Here $p_0$ is the probability 
that the PE chooses a cell $c$ 
so that $|W(c)|=0$;
\\
~~~~~~(ii) with probability $p_1 =4(m-2)/m^2$,
the PE must check the $h(C')$ of one of its four neighbors $C'$
before making the update.
The $C'$ is chosen uniformly at random among the four possibilities.
If $h(C') ~\geq~ h(C)$,
then $h(C)$ gets an increment according to \eqref{hC};
otherwise, $h(C)$ is not updated.
Here $p_1$ is the probability that PE will choose
a cell $c$ in an edge but not in a corner, so that $|W(c)|=1$
\\
~~~~~~(iii) with the remaining probability $p_2=4/m^2$,
the PE checks $h(C')$ and $h(C'')$
of two of its adjacent neighbors
(for example in Figure~\ref{fig:aggr}, neighbors PE2 and PE3
can be involved in the computation for PE1).
The two neighbors are chosen uniformly at random
from the four possibilities.
Again, if both
$h(C')  \geq  h(C)$
and
$h (C'') \geq  h(C)$,
then $h(C)$ gets an increment according to \eqref{hC};
otherwise, $h(C)$ is not updated.
Here $p_2$ is the probability 
to choose a cell $c$ in a corner,
so that $|W(c)|=2$.

\begin{figure}
\centering
\includegraphics*[width=5.8in]{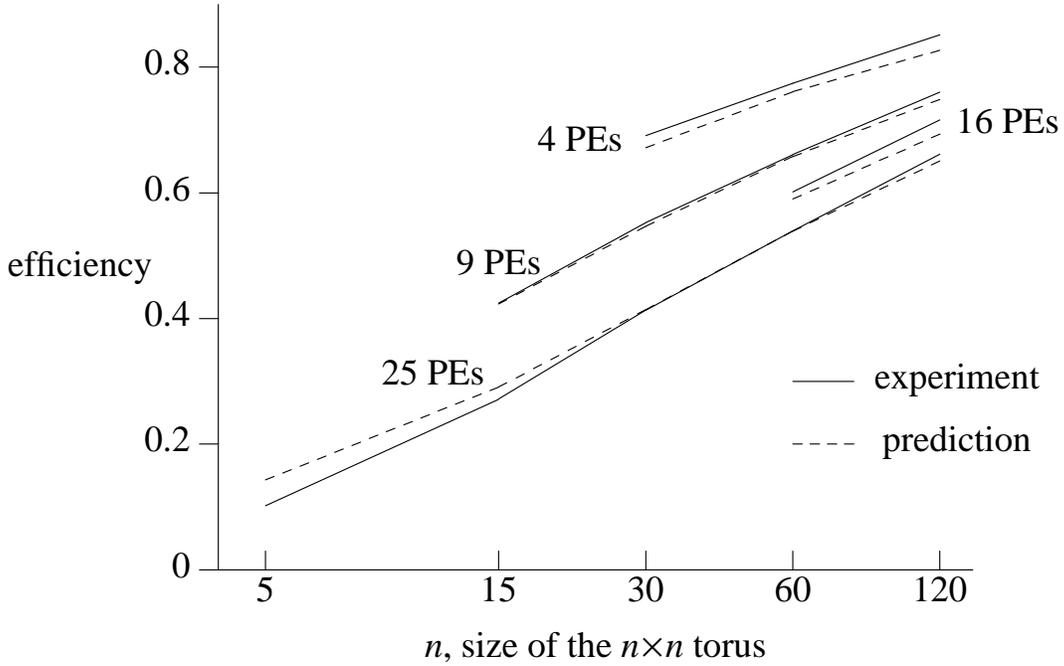}
\caption{Performance of the Ising model simulation. Many-cells-per-one-PE case}
\label{fig:perfmt1}
\end{figure}

As in the previous case,
this simplified model simulates
a possible but not obligatory synchronous timing arrangement for
executing the real asynchronous algorithm.
Figure~\ref{fig:perfmt1} shows excellent agreement between
actual and predicted performances
for the aggregated Ising model.
The efficiency presented in Figure~\ref{fig:perfmt1} is computed as

\begin{equation}
\label{effi}
{\rm efficiency}=
\frac{\rm serial~execution~time} {{\rm number~of~PEs} \times {\rm parallel~execution~time}}
\end{equation}

The parallel speed-up can be found as 
efficiency$~\times~$number~of~PEs.
For 25 PEs simulating a 120$\times$120 Ising model,
efficiency is 0.66;
hence, the speed-up is greater than 16.
For the currently unavailable sizes,
when $10^4$ PEs
simulate a $10^4 \times 10^4$ array,
the simplified model predicts
an efficiency of about 0.8 and a speed-up of about 8000.

\begin{figure}
\centering
\includegraphics*[width=5.8in]{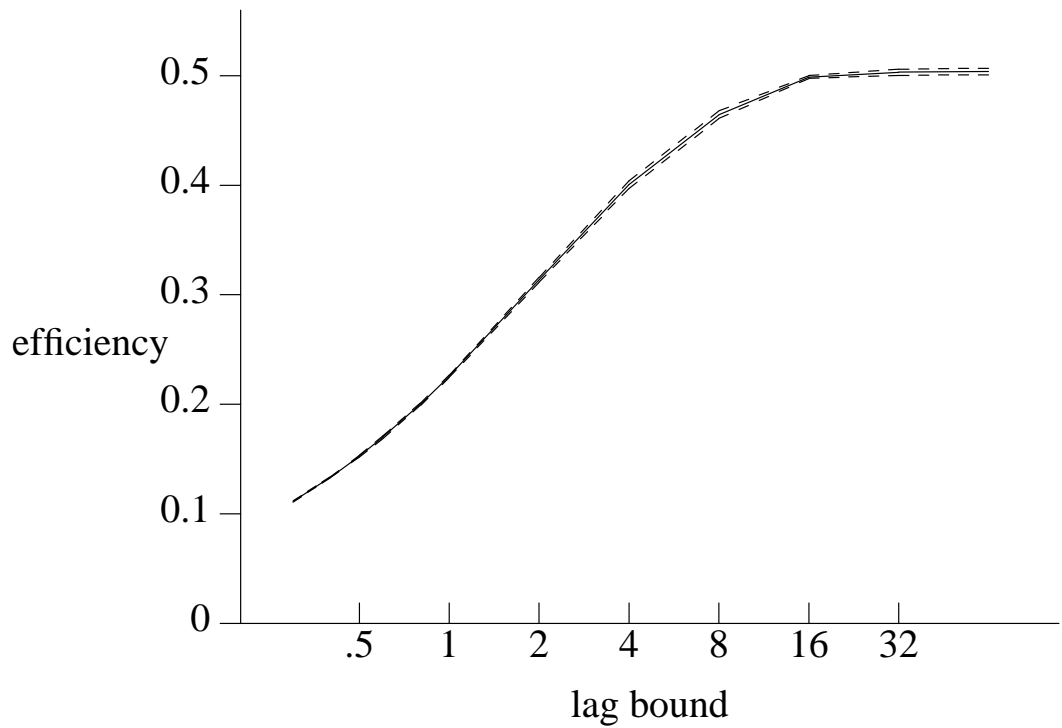}
\caption{Efficiency degradation caused by bounded lag}
\label{fig:perfbl}
\end{figure}

In the experiments reported above,
the lag between the local times of any two PEs
was not restricted.
As discussed in Section~\ref{sec:algo},
an upper bound on the lag 
might result from the necessity
to produce the output.
To see how the bound
affects the efficiency,
one experiment reported in Figure~\ref{fig:perfmt1},
is repeated with various finite 
values of the lag bound.
In this experiment, 
an $n \times n$ array is simulated
and
one PE carries an $m \times m$ subarray,
where $n=384$ and $m=12$.
The results are presented in Figure~\ref{fig:perfbl}.

In Figure~\ref{fig:perfbl},
the unit of measure for a lag is the expectation of
time intervals between consecutive arrivals for a cell.
For lag bounds greater than 16, 
degradation of efficiency is almost unnoticeable,
when
compared with the base experiment where lag$= \infty$.
Substantial degradation starts at about 8;
for the unity lag bound,
the efficiency is about half that of the base experiment.
However, even for lag bound 0.3, the simulation remains practical,
with an efficiency of about 0.1;
since 1024 PEs execute the task, 
this efficiency means a speed-up of more than 100.

\section{Conclusion}\label{sec:concl}
\hspace*{\parindent} 
This paper demonstrates an efficient parallel method
for simulating asynchronous cellular arrays.
The algorithms are quite simple and easily implementable
on appropriate hardware.
In particular, each algorithm
presented in the paper
can be implemented on a general purpose
asynchronous parallel computer,
such as the currently available bus machines with shared memory.
The speed of such implementation
depends on the speed of PEs 
and the efficiency of the communication system.
A crucial condition for success in such implementation
is the availability of a good parallel generator 
of pseudo-random numbers.
To assure reproducibility,
each PE should have its own reproducible
pseudo-random sequence.

The proposed algorithms present
a number of challenging mathematical problems,
for example, the problem of 
proving that efficiency tends 
to a positive limit when the number of PEs increases
to infinity.
\\

{\bf Acknowledgments}.
\\
I acknowledge the personnel
of the Thinking Machine Corp. for their kind invitation,
and help in debugging and running the parallel *LISP
program on one of their computers.
Particularly, the help of 
Mr. Gary Rancourt and Mr. Bernie Murray was invaluable.
Also, I thank Andrew T. Ogielski and Malvin H. Kalos 
for stimulating discussions, 
Debasis Mitra for a helpful explanation of a topic in Markov chains,
and Brigid Moynahan for carefully reading the text.


\newpage
{\bf APPENDIX: a working code of Ising simulation}
\\
\\
C language program for the BALANCE parallel computer;
the code is used for timing only and contains no i/o;
the code of the pseudo-random number generator 
is not included 
\begin{verbatim}
#include <pp.h>
#include <math.h>
#include <sys/tmp_ctl.h>

#define SHARED_MEM_SIZE (sizeof(double)*10000)
#define END_TIME 1000.
#define A 24       /* side of small square a PE takes care of*/
#define M 5        /* number of PEs along a side of the big square*/

shared int nPEs = M*M, spin[M*A][M*A];
shared float time[M][M];  /*local times on subarrays*/
shared float prob[10];  /* probabilities of state change */
shared float J = 1., H = 0.;     /* Energy= -J sum spin spin' - H sum spin */
shared float T = 1.;                /* Temperature */
shared int ato2 = A*A;
shared int am = A*M;

main()
{
    int i,j,child_id, my_spin, sum_nei, index, bit;
    float d_E, x;
    double frand();

/* compute flip probabilities */
    for (i = 0; i < 5 ; i++)
        for (j = 0; j < 2; j++)
          {index = i + 5*j;    /* index = 0,1,...,9 */ 
           my_spin = 2*j - 1;
           sum_nei = 2*i - 4;
           d_E = 2.*(J * my_spin * sum_nei + H * my_spin);
           x = exp(-d_E/T);
           prob[index] = x/(1.+x);
   /*      printf("prob[%d]=%f\n",index,prob[index]);  */
          };

/* initialize local times */
    for (i = 0; i < M; i++)
        for (j = 0; j < M; j++)
            time[i][j]=0.;

/* initialize spins at random, in seedran(seed,b), b is dummy*/
    seedran(31234,1);
    for (i = 0; i < M*A; i++)
        for (j = 0; j < M*A; j++) {
            bit = 2*frand(1);                /* bit becomes 0 or 1 */
            spin[i][j] = 2*bit - 1;          /* spin becomes -1 or 1 */
   /*       printf("spin[%d][%d]=%d\n",i,j,spin[i][j]);      */
    };

    /* in the following loop single PE spawns nPEs other PEs for concurrent
       execution. Each child PE would execute subroutine work(my_id) with its
       own argument my_id. */

    for (child_id = 0; child_id < nPEs; child_id++)
          if (fork() == 0) {
              tmp_affinity(child_id);     /* fixing a PE for process child_id */
              work(child_id);              /* starting a child PE process */
              exit(0);
          }


    /* in the following loop the parent PE awaits termination of each child PE
       then terminates itself */
    for (child_id = 0; child_id < nPEs; child_id++) wait(0);
    exit(0);
}

work(my_id)
int my_id;
{
  int i,j;
  int coord, var;
  int x,y,my_i,my_j,sum_nei, nei_i,nei_j;
  int  up_i, down_i, left_j, right_j;
  int i_base, j_base;
  int index;
  double frand(); 
  double r;
  double end_time;

  end_time = END_TIME*A*A;
                               /*normalizing time scale for multiprocessor execution*/

  my_i = my_id%M;      /*PE my_id carries small square (my_i,my_j)*/
  i_base = my_i*A;     
  up_i = (my_i + 1)%M; 
  down_i = (my_i + M - 1)%M;

  my_j = (my_id-my_i)/M;
  j_base = my_j*A;        
  left_j = (my_j + M - 1)%M;
  right_j = (my_j + 1)%M;

  seedran(my_id*my_id*my_id,my_id);  
 /*PE my_id has its own copy of pseudo-random number generator and initializes it 
   using seedran(seed,my_id) with unique seed=my_id*my_id*my_id */

  while(time[my_i][my_j] < end_time) 
  {
    r = frand(my_id); 
      /*PE my_id obtains next pseudo-random number from its own sequence*/
    x = r*A;               
    y = (r*A-x)*A;  
       /*pick a random cell with internal address (x,y) within the A*A square*/

/*compute sum of neighboring spins*/
    sum_nei = 0;          
    for (coord = 0;  coord < 2; coord += 1)
        for (var = -1;  var < 2; var += 2)
    {
          nei_i = x;
          nei_j = y;
          if(coord == 0) nei_i += var;
          if(coord == 1) nei_j += var;

          if(0 <= nei_i && nei_i < A && 0 <= nei_j && nei_j < A) 
          {
             nei_i += i_base;
             nei_j += j_base;
          }
          else 
          {
       /* 4 possible reasons to wait for a neighboring PE */
            if(-1 == nei_i) while (time[down_i][my_j]  < time[my_i][my_j]) ;
            if(-1 == nei_j) while (time[my_i][left_j]  < time[my_i][my_j]) ;
            if(nei_i == A)  while (time[up_i][my_j]    < time[my_i][my_j]) ;
            if(nei_j == A)  while (time[my_i][right_j] < time[my_i][my_j]) ;

            nei_i = (nei_i+i_base+am)%am;
            nei_j = (nei_j+j_base+am)%am;   
          };
          sum_nei += spin[nei_i][nei_j];
    };

/*recover index*/
    index = (sum_nei + 4)/2 + 5*(spin[x+i_base][y+j_base] + 1)/2;

    r = frand(my_id); 

    if(r < prob[index]) 
      spin[x+i_base][y+j_base] *= -1;
    else /* printf(": NO flip\n") */ ;

    r = frand(my_id); 
    time[my_i][my_j] += -log(r);
  };
}
\end{verbatim}
\end{document}